\documentclass[twocolumn,prl,floatfix]{revtex4}

\usepackage{color}
\usepackage{epsfig}

\begin{document}

\title{Velocity selection problem for combined motion 
of melting and solidification fronts}
\author{Efim A. Brener and D. E. Temkin}
\affiliation{Institut f\"ur Festk\"orperforschung, Forschungszentrum J\"ulich,
D-52425 J\"ulich, Germany} 

\begin{abstract}
We discuss a free boundary problem for two moving solid-liquid 
interfaces that strongly interact 
via the diffusion field in the liquid layer between them. This problem arises 
in the context of liquid film migration (LFM) during the partial melting of solid alloys.
In the LFM mechanism the system chooses 
a  more efficient kinetic path which is controlled by  diffusion in the liquid film, 
whereas the  process with only one melting front would be controlled
 by the very slow diffusion in the mother solid phase.
The relatively weak coherency strain energy 
is the effective driving force for LFM.
As in the classical dendritic growth problems, also in this case  
an exact family of steady-state solutions  
with two parabolic fronts and an arbitrary velocity exists  
if capillary effects are neglected
 \cite{temkin2005}.  
We develop a velocity selection theory for this problem,
including  anisotropic surface tension effects.
The strong diffusion interaction and coherency strain 
effects in the solid near the melting
front  lead to substantial changes compared to classical dendritic growth. 
\end{abstract}

\maketitle

The early observation of liquid film migration (LFM) were made
during sintering in the presence  of liquid phase \cite{yoon1} 
or during partial melting of alloys \cite{musch1}  
(see  \cite{yoon2} for a review).  Nowadays LFM
is a well established phenomenon of great practical importance.  
In LFM one  crystal is melted 
and another one is solidified.  Both solid-liquid interfaces 
move  together with the same velocity. In the investigated 
alloys systems the migration velocity is of the order of $10^{-6}-10^{-5}$ 
cm/s and it is controlled  by the solute diffusion through 
a thin liquid layer between the two interfaces \cite{4}. 
The migration 
velocity is much smaller than the characteristic velocity of 
atomic kinetics at the interfaces. Therefore,  both solids should at the 
interfaces  be  locally in thermodynamic equilibrium with the  
liquid phase. On the other hand,  these local equilibrium 
states should be different for the two interfaces 
to provide the driving force for the process. 
 It is by now well accepted  (see, for example, \cite{yoon2,4}) 
that the difference of the equilibrium states at the melting and solidification fronts 
is due to the coherency strain energy,  
important only at the melting front because of the sharp concentration profile  
ahead the moving melting front (diffusion in the solid phase is very slow and the 
corresponding diffusion length is very small). 
Thus, the liquid composition
at the melting front, which depends on the coherency strain energy 
and on the curvature of the front, differs from the liquid composition 
at the unstressed and curved solidification front. This leads to the 
necessary gradient of the concentration across the liquid film and the process is 
controlled by the diffusion in the film.

If only the melting front existed,  
the melting process would be controlled by the very slow diffusion 
in the mother solid phase 
and  elastic effects would be irrelevant. In the LFM mechanism the system chooses 
a  more efficient kinetic path  which 
 is controlled by the much faster diffusion in the liquid film. However, 
in this case 
the relatively weak coherency strain energy is  involved as  effective driving 
force for this process. In this respect the LFM mechanism is similar to  other 
well-known phenomena such as diffusion induced grain boundary migration and 
cellular precipitation \cite{yoon2}. In these processes a relatively fast diffusion 
along the grain boundaries controls the kinetics and the coherency strain energy  
plays also a controlling role.

Thus, a theoretical description of LFM requires the solution
of a free boundary problem for two combined moving solid-liquid 
interfaces with a liquid film in between. In Ref.\cite{temkin2005}  
this problem was considered for simplified boundary conditions: 
the temperature and the chemical composition along each interface 
were kept constant. Their values are different for the melting and 
solidification fronts and differ from those far from the migrating 
liquid film. This means that any capillary, kinetic and crystallographic 
effects at the interfaces were neglected. It was found that under these simplified 
boundary conditions two co-focal parabolic fronts can move together 
with the same velocity. The situation is rather similar to a steady-state  motion 
of one parabolic solidification front into a supercooled melt \cite{ivan1,ivan2} 
or one parabolic melting front into a superheated solid. 
In this approximation the Peclet numbers were found, but the steady-state velocity 
remained undetermined at that stage. Thus, the problem of velocity selection arises.

Solvability theory has been very successful in predicting certain properties of 
 pattern selecting in dendritic growth and a number of related phenomena 
(see, for example, \cite{saito96,kessler88,brener91}).
In the two-dimensional dendritic case, the basic approach is as follows. 
One attempts to model the dendritic tip by a needle crystal, that is, a shape-preserving 
steady-state growth shape which is a solution of the equation of motion governing 
diffusion in the neighborhood of a solidification front. This needle crystal 
is assumed to be close in shape to the parabolic Ivantsov solution. If anisotropic 
capillary effects are included,  a single dynamically stable solution is found 
for any external growth conditions. This theory has been extended to the 
three-dimensional case \cite{ben93,brener93}. We note that capillarity is 
a singular perturbation and the anisotropy of the surface energy is a prerequisite 
for the existence of the solution. 

The main purpose of this Letter is to develop a velocity selection theory for LFM, 
including in the consideration anisotropic surface tension effects.
We note that this is not a just 
routine extension of the existing  theory because the diffusion interaction between 
two interfaces changes the problem substantially. 

\begin{figure}
\begin{center}
\epsfig{file=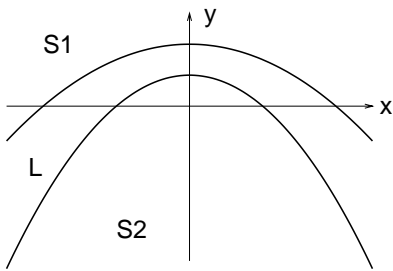, width=6cm}
\caption{Schematic presentation of two  moving nearly parabolic fronts;
$S1$ and $S2$ are the melting and growing solids, and $L$ is the liquid film.}
\end{center}
\end{figure}

We discuss  the two-dimensional problem of the steady-state motion of a 
thin liquid film during the process of 
isothermal melting of a binary alloy, see Fig.1. 
We assume that the diffusion in 
the solid phases is very slow and the concentration $c$ in the liquid film obeys the 
Laplace equation. We introduce the normalized concentration $C=(c-c_L)/(c_L-c_S)$
with $c_L$ and $c_S$ being the  liquidus and solidus 
concentrations of the equilibrium phase diagram at a given temperature.
Then the equilibrium concentration and the mass balance conditions
at the solidification front read
\begin{equation}
C=d_2K_2, \qquad  V_n=-D\partial C/\partial \bf{n}.
\end{equation}
At the melting front the equilibrium concentration is changed 
by the presence of the elastic coherency strain energy \cite{yoon2} 
and  also the diffusional flux changes 
 because in the solid ahead of the melting front 
the concentration is $c_0$ which is different from $c_S$:
\begin{equation}
C=-b\Delta^2-d_1K_1, \qquad  V_n(1-\Delta)=-D\partial C/\partial \bf{n}.
\end{equation}
Here $V_n$ is the normal velocity; 
$D$ is the diffusion coefficient in the liquid film; $K$ is the curvature assumed 
to be negative for the interfaces in Fig.1; $\Delta=(c_0-c_S)/(c_L-c_S)$ is the 
dimensionless driving force; $b=Y\Omega(da/dc)^2/a^2f_L''$ 
is the dimensionless constant which describes the 
coherency strain energy \cite{4}, $\Omega$ is the atomic volume, 
$Y$ is the bulk elastic modulus,  $a$ is the atomic constant,
$f_L(c)$ is the free energy of the liquid phase per atom,
$f_L''$ is the second derivative of $f_L(c)$ at $c=c_L$; 
$d_i$ are the anisotropic chemical capillary lengths:
\begin{equation}
 d_i(\theta)=d_0[1-\alpha\cos 4(\theta-\theta_i)],\quad 
d_0=\gamma\Omega / f_L''(c_L-c_S)^2
\label{anis}
\end{equation}
with the isotropic part of the surface energy $\gamma$ and 
the anisotropy parameter $\alpha<<1$; 
$\theta$ is the angle between the normal to the interface $\bf{n}$
and the direction of motion,
$\theta_i$ is the direction of the minimum of $d_i(\theta)$ for each of the 
interfaces.

We measure all lengths in the units of the radius of curvature of the Ivantsov parabolic
 solidification front, $R_2$. Introducing  a parabolic coordinates system,
\begin{equation}
y=(\eta^2-\beta^2)/2, \quad \quad x=\eta\beta,
\end{equation} 
we look for a solution of the Laplace equation for the concentration 
field $C(\eta,\beta)$
in the 
liquid phase with the following boundary conditions: 
$$C(\eta=\eta_2(\beta),\beta)=d_2(\beta)K_2(\beta),$$
\begin{equation}
C(\eta=\eta_1(\beta),\beta)=-b\Delta^2-d_1(\beta)K_1(\beta).
\end{equation}
The mass balance conditions at the interfaces are:
$$
2P(\eta_2(\beta)+\beta\eta'_2(\beta))=
-(\partial C/\partial\eta-\eta'_2\partial C/\partial\beta)
$$
\begin{equation}
2P(1-\Delta)(\eta_1(\beta)+\beta\eta'_1(\beta))
=-(\partial C/\partial\eta-\eta'_1\partial C/\partial\beta)
\end{equation}
where the Peclet number is $P=VR_2/2D$ and $V$ is the steady-state velocity in the 
$y$-direction. 
 We present the two moving interfaces in the form,
\begin{equation}
\eta_2=1+\epsilon_2(\beta),
\quad \quad
 \eta_1=\eta_0+\epsilon_1(\beta)
\end{equation}
and the concentration field in the form,
\begin{equation}
C(\eta,\beta)=C_0(\eta)+u(\eta,\beta),
\end{equation}
where $C_0(\eta)=-b\Delta^2(\eta-1)/(\eta_0-1)$ is the solution without 
surface tension ($u=\epsilon_1=\epsilon_2$=0) which satisfies 
the boundary conditions: $C_0(1)=0$ and $C_0(\eta_0)=-b\Delta^2$. The balance equations
at the interfaces,
$2P=2P(1-\Delta)\eta_0=-\partial C_0/\partial\eta$ lead to the relations for the 
Peclet number and $\eta_0$: 
\begin{equation}
 P=\frac{VR_2}{2D}=\frac{b\Delta^2}{2(\eta_0-1)},
\quad  \eta_0=\sqrt{\frac{R_1}{R_2}}=\frac{1}{1-\Delta}.
\label{peclet}
\end{equation}
 These relations can also be obtained in 
the proper limit of Eqs.(27)-(28) of Ref. \cite{temkin2005} and give the expressions 
 for the radii of curvature for the two interfaces, $R_1$ and $R_2$, for a given 
velocity $V$ which remains undetermined at this stage.
 
The surface tension plays a  
crucial role in the velocity selection problem. 
In order to find the small corrections to these 
solutions due to the small (but singular) 
surface tension effects we  should first  find the 
small correction $u$ to the diffusional 
field in the linear approximation with respect to $\epsilon$ and $dK$ (see, for example,
\cite{brener91}). In this approximation, the field $u$ satisfies the Laplace equation 
and  the boundary conditions:
$$
u(\eta=1,\beta)=\Psi_2(\beta)=2P\epsilon_2+d_2K_2,   
$$
\begin{equation}
u(\eta=\eta_0,\beta)=\Psi_1(\beta)=2P\epsilon_1-d_1K_1.
\label{bounde}
\end{equation}
 The balance equations read:
$$
2P(\epsilon_2+\beta\epsilon'_2)=-\partial u/\partial\eta|_{\eta=1}
$$
\begin{equation}
 2P(1-\Delta)(\epsilon_1+\beta\epsilon'_1)=-\partial u/\partial\eta|_{\eta=\eta_0}.
\label{fluxe}
\end{equation}
The Laplace equation with the boundary conditions, Eqs.(\ref{bounde}), is easily 
solved by  Fourier transform, with respect to the variable $\beta$, which we define as
$$
\overline{g(\lambda)}=
\frac{1}{\sqrt{2\pi}}\int_{-\infty}^{\infty}d\beta\exp(-i\lambda\beta)g(\beta)
$$
$$
g(\beta)=\frac{1}{\sqrt{2\pi}}\int_{-\infty}^{\infty}
d\lambda\exp(i\lambda\beta)\overline{g(\lambda)}.
$$
Then,  
\begin{equation}
\overline{u(\eta,\lambda)}=
\frac{\overline{\Psi_1}
\sinh [\lambda(\eta-1)]}{\sinh [\lambda(\eta_0-1)]}-
\frac{\overline{\Psi_2}
\sinh [\lambda(\eta-\eta_0)]}{\sinh [\lambda(\eta_0-1)]}.
\end{equation}
Finally, Eqs.(\ref{fluxe}) read
$$
2P\frac{d\overline{\epsilon_2}}{d\lambda}=
\frac{\overline{\Psi_1}}{\sinh[\lambda(\eta_0-1)]}-
\frac{\overline{\Psi_2}\cosh[\lambda(\eta_0-1)]}{\sinh[\lambda(\eta_0-1)]}$$
\begin{equation}
2P(1-\Delta)\frac{d\overline{\epsilon_1}}{d\lambda}=
\frac{\overline{\Psi_1}\cosh[\lambda(\eta_0-1)]}{\sinh[\lambda(\eta_0-1)]}
-\frac{\overline{\Psi_2}}{\sinh[\lambda(\eta_0-1)]}.
\end{equation}
Eliminating $\overline{\epsilon_1}$ from the first of these two equations,
$$
2P\overline{\epsilon_1}= 2P(\overline{\epsilon_2})'\sinh[\lambda(\eta_0-1)]
+2P\overline{\epsilon_2}\cosh[\lambda(\eta_0-1)]+$$
\begin{equation}
\overline{d_1K_1}+
\overline{d_2K_2}\cosh[\lambda(\eta_0-1)],
\end{equation}
we find from the second one,
$$2P(\overline{\epsilon_2})''-2P \overline{\epsilon_2}=\overline{d_2K_2} $$
\begin{equation}
-\left[\frac{(\overline{d_1K_1})'}
{\sinh[\lambda(\eta_0-1)]}
+(\overline{d_2K_2})'\frac{\cosh[\lambda(\eta_0-1)]}
{\sinh[\lambda(\eta_0-1)]}\right ]
\label{fur2}
\end{equation}
Here $'$ denotes derivatives with respect to $\lambda$.

These equations are convenient for the further analysis. 
We are interested in the motion of the thin film, which means that $(\eta_0-1)\approx 
\Delta << 1$. In this case we can expand the hyperbolic functions 
for small values of their argument almost everywhere. This leads to $\epsilon_1\approx
\epsilon_2$, $K_1\approx K_2$ and then, from Eq.(\ref{fur2}) we find  in  
direct space:
 \begin{equation}  
 \frac{d}{d\beta}\left[(1+\beta^2)\epsilon_2\right]=\frac{d_1+d_2}{2P\Delta}\beta K_2
\label{local} 
\end{equation}   
A few remarks are in order.
First of all, this is somewhat the expected result 
because for the thin films the resulting 
equation should be local and, in principal, can be derived in the framework of boundary 
layer techniques. However,  the used approximation for small 
arguments of hyperbolic functions breaks down in the small vicinity 
(of the order of $\Delta$) near the 
singular points in the complex plane, $\beta=\pm i$. We will discuss this point 
later. 
Finally,  we can return to  Cartesian coordinates because the 
expression for the curvature is much simpler in this representation. 
In the linear approximation $(1+\beta^2)\epsilon_2\approx \zeta_2(x)$ where $\zeta_2(x)$ 
is the correction to the parabolic solution in the Cartesian coordinates.
Then Eq.(\ref{local}) reads:
  \begin{equation}  
 \frac{d\zeta_2}{dx}=\frac{d_1+d_2}{2P\Delta}x K_2
\label{localc} 
\end{equation}
One finds the first regular corrections to the parabolic 
shape  by replacing $d_1=d_2=d_0$ and 
$K_2=-1/[R_2(1+x^2)^{3/2}]$ for the parabola:
\begin{equation}
\zeta_2(x)= \sigma/[\Delta (1+x^2)^{1/2}], \quad \Delta<<1,
\label{reg}
\end{equation}
where $\sigma=d_0/(PR_2)$ is the usual stability parameter which appears 
in such  problems.

In the vicinity of the  point $x=-i$ where the curvature becomes singular and anisotropy,
Eq.(\ref{anis}),
becomes  important we rescale the variables \cite{brener91}:
\begin{equation}
x=-i(1-\sqrt{\alpha}z), \quad \zeta_2=\alpha F,  \quad \tau=z+dF/dz,
\label{resc}
\end{equation} 
with a small anisotropy parameter, $\alpha<<1$.
Then Eq.(\ref{localc}) reads
\begin{equation}
\frac{dF}{dz}=- \frac{\sigma}{\Delta\alpha^{5/4}}A(\tau)\frac{1+d^2F/dz^2}{(2\tau)^{3/2}}
\label{sing1}
\end{equation}
where the anisotropy factor is $A(\tau)=1 -[\exp(-4i\theta_1)+\exp(-4i\theta_2)]/\tau^2$.
This equation is similar to the limit of the kinetic dendrite in 
Eq.(9.1) of Ref.\cite{brener91}.  This similarity is purely formal because the real 
interface kinetic effects 
are not included in our description. Instead,  the strong diffusion interaction leads 
to such a selection problem. The same arguments as in \cite{brener91} give 
the selection condition
\begin{equation} 
\sigma=\sigma^{\star}\sim \Delta\alpha^{5/4}, \qquad \Delta<< \sqrt{\alpha}.
\label{sel1}
\end{equation}
Moreover, the selected growth direction lies in between  
 the directions of  the minima 
of $d_1(\theta)$ and $d_2(\theta)$ for the two interfaces, 
$\theta_1=-\theta_2$, because the coefficient
 appearing in $A(\tau)$ must be real \cite{brener91}. 

While the regular correction given by Eq.(\ref{reg}) is valid for $\Delta <<1$, the 
restriction for the validity of inner Eq.(\ref{sing1}) 
and for the scaling relation, Eq.(\ref{sel1}), 
is much stronger, $\Delta << \sqrt{\alpha}<<1$.
In the opposite limit, $\Delta >> \sqrt{\alpha}$, we can derive another local inner 
equation where the important scale $|x+i|\sim\sqrt{\alpha}$
is much smaller then $\Delta$. The argument of the 
hyperbolic functions becomes large and we can neglect the term $K_1$ in Eq.(\ref{fur2}).
Then, in the vicinity of the singular point $x=-i$ in the complex plane,  
this equation reads:
\begin{equation}
\zeta_2(x)=-d_2K_2/P,
\end{equation}
which is the usual inner equation for a single dendrite in the one-sided model 
\cite{brener91}. With the same rescaling as in  Eq.(\ref{resc}), we find    
\begin{equation}
F=\frac{\sigma}{\alpha^{7/4}}A(\tau)\frac{1+d^2F/dz^2}{(2\tau)^{3/2}},
\label{sing2}
\end{equation}
where the anisotropy factor is $A(\tau)=1 -2\exp(-4i\theta_2)/\tau^2$.
This corresponds to the selection condition
\begin{equation} 
\sigma=\sigma^{\star}\sim \alpha^{7/4}, \qquad \Delta>>\sqrt{\alpha}
\label{sel2}
\end{equation}
and the selected growth direction corresponds to the minimum of $d_2(\theta)$, 
$\theta_2=0$. Whereas  Eq.(\ref{sing2}) and 
the selection condition Eq. (\ref{sel2}) are valid not only for small $\Delta$ 
\cite{brener91},  
the regular correction in the form of Eq.(\ref{reg}) requires 
$\Delta<<1$. 

The selection conditions, Eqs.(\ref{sel1},\ref{sel2}), together with the relations 
for the Peclet numbers, Eq.(\ref{peclet}), solve the posed problem of pattern formation 
for two combined nearly parabolic fronts:
\begin{equation}
V=\frac{2D}{d_0}P^2(\Delta)\sigma^{\star}
\label{vel}
\end{equation}
and 
\begin{equation}
R_2=d_0/[P(\Delta)\sigma^{\star}], \quad R_1=R_2/(1-\Delta)^2.
\label{R}   
\end{equation}

While these results are formally similar to the free dendritic case, 
the selected stability
parameter $\sigma^{\star}$ scales as $\alpha^{7/4}$ only for $\Delta>>\sqrt{\alpha}$.
Otherwise, the other scaling relation, Eq.(\ref{sel1}), holds. In principal, it could 
be conceivable that the strong diffusional interaction between the front leads to 
the selection even without anisotropy. However, our analysis of the inner equation 
does not support this hypothesis. The Peclet number is 
$P(\Delta)\sim b\Delta$ which reflects the fact that the coherency 
strain energy plays a crucial role  
in the LFM mechanism. This parameter $b$ is usually small but the melting process by 
the LFM mechanism is controlled by  the fast diffusion in the liquid, whereas the  
the process with only one melting front would be controlled
 by the very slow diffusion in the mother solid phase. 
As we already noted, with the help of 
the LFM mechanism the system chooses  a more efficient kinetic path to relax to 
the equilibrium state.

Finally, we estimate the velocity $V$ from Eq.(\ref{vel}) and the thickness of the film
$R_2\Delta$  from Eq.(\ref{R})
using  characteristic values of the parameters: $D\sim 10^{-5}$cm$^2$/s, 
$d_0\sim 10^{-7}$cm, $b\sim 0.05$,  $\Delta\sim 0.05$ and $\sigma^{\star}\sim 10^{-2}$. 
This leads to $V\sim 10^{-5}$cm/s and $R_2\Delta\sim 10^{-4}$cm, 
which qualitatively agree with typical values in LFM experiments.

In conclusion, we developed a selection theory for the process of liquid film migration 
where the strong diffusion interaction between melting and solidification fronts plays
a crucial role. This process is very important in practical applications, in particular 
during sintering  in the presence 
of the liquid phase \cite{yoon1,yoon2}. However, despite  its practical importance, 
experimental investigations of this process are  far from the level of accuracy 
of the model experiments in classical dendritic growth. 
Our approach extends selection theory developed for dendritic growth to LFM and  
we hope that our results will stimulate  further theoretical and 
experimental investigations in this very interesting field.
From the theoretical side it would be a challenge to attack this problem   by a direct 
numerical approach, for example, by means of the phase-field model.

We acknowledge the support by the Deutsche Forschungsgemeinschaft under 
project SPP 1120.


\begin{thebibliography}{99}
\bibitem{yoon1} D. N. Yoon, and W. J. Hupmann, Acta Metall {\bf 27}, 973 (1979). 
\bibitem{musch1} T. Muschik T, W. A. Kaysser,  and T. Hehenkamp, 
Acta Metall {\bf 37}, 603 (1989).
\bibitem{yoon2} D. N. Yoon, Int. Mater. Rev. {\bf 40}, (1995). 
\bibitem{4} D. N. Yoon, J. W. Cahn,  C.A.  Handwerker,  J. E.Blendell,
  and Y. J. Baik, In: Interface Migration and Control of Microstrucutres.
  Am Soc. Metals. Park. Ohio (1985), pp. 19-31. 
\bibitem{temkin2005} D.E. Temkin , to be published in Acta Materialia.
\bibitem{ivan1} G. P. Ivantsov, Dokl. Akad. Nauk SSSR {\bf 58}, 567 (1947).
\bibitem{ivan2} G. P. Ivantsov, Dokl. Akad. Nauk SSSR {\bf 83}, 573 (1952).
\bibitem{saito96} Y. Saito, Statistical Physics of Crystal Growth, World
  Scientific Publishing, Singapore, 1996.
\bibitem{kessler88} D. Kessler, J. Koplik, and H. Levine, 
Adv. Phys. {\bf{37}}, 255 (1988).
\bibitem{brener91}  E. A. Brener, and V. I. Mel'nikov, Adv. Phys. {\bf 40} 53 (1991).
\bibitem{ben93} M. Ben Amar, and E.A. Brener,  Phys. Rev. Lett {\bf{71}}, 589 (1993).
\bibitem{brener93} E.A. Brener,  Phys. Rev. Lett. {\bf{71}}, 3653 (1993).
\end{thebibliography}
\end{document}